\documentstyle[aps,amssymb,12pt]{revtex}

\begin{document}
\title{Spin-Dependent Transport Through An Interacting Quantum Dot}
\author{Ping Zhang$^1$, Qi-Kun Xue$^1$, Yu-Peng Wang$^1$, X.C. Xie$^{1,2}$}
\address{$^1$International Center of Quantum Structure and State Key Laboratory for\\
Surface Physics, Institute of Physics, The Chinese Academy of Sciences,\\
Beijing 100080, P.R. China\\
$^2$Department of Physics, Oklahoma State University, Stillwater, OK 74078 }
\maketitle

\begin{abstract}
We study the nonequilibrium spin transport through a quantum dot containing
two spin levels coupled to the magnetic electrodes. A formula for the
spin-dependent current is obtained and is applied to discuss the linear
conductance and magnetoresistance in the interacting regime, where the
so-called Kondo effect arises. We show that the Kondo resonance and the
correlation-induced spin splitting of the dot levels may be systematically
controlled by internal magnetization in the electrodes. As a result, when
the electrodes are in parallel magnetic configuration, the linear
conductance is characterized by two spin-resolved peaks. Furthermore, the
presence of the spin-flip process in the dot splits the Kondo resonance into
three peaks.%
\newline%
PACS number(s): 73.23.-b, 73.63.-b, 75.25.+z%
\newline%
\end{abstract}

Spin-polarized transport in magnetic nanostructures, in particular,
single-electron tunneling in ferromagnetic (F) double junctions, has become
a very active area of research, mainly because of its possible applications
in information storage and processing devices\cite{Prinz}. In these
junctions, the transport properties depend on the relative orientation of
the magnetic moments of external electrodes. When the magnetic moments of
the electrodes are antiparallel, the tunnel resistance increases. This is
called as the tunnel magnetoresistance (TMR). When the central grain in
double junctions is small enough to form a quantum dot (QD), the effects of
the discrete quantized energy levels as well as Coulomb blockade become
significant, which has been considered theoretically and experimentally in
the sequential tunneling regime\cite
{Ono,Sch,Tak,Wang,Bar,Mar,Bul1,Bar2,Bul2,Rud}.

In the low temperature regime, a more subtle effect of large charging
energies is the creation of new states of many-body character at the Fermi
level by the Kondo effect\cite{Hew}. For a QD coupled to the normal (N)
electrodes, the Kondo effect is well understood in and out of equilibrium%
\cite{Gla,Ng1,Mei1,Win}. It is a consequence of a special kind of high-order
tunneling process in which the electron inside the QD tunnels out followed
by an electron with opposite spin tunneling into the QD. The whole system
forms a spin singlet state and the net magnetic moment in the QD is zero.
For the F-QD-F system, a very important question is, what are the
consequence and characteristics of the Kondo effect when the magnetic
moments of the electrodes are taken into account?

In this Letter we have studied spin-dependent transport in an interacting QD
coupled to two magnetic electrodes as shown schematically in Fig. 1.
Different from the conventional Kondo problem in a N-QD-N system, here the
characteristics induced by the strong electronic correlation is sensitive to
the relative orientation of magnetization between the two electrodes,
namely, the parallel and antiparallel configurations as shown in Fig. 1.

The model Hamiltonian for the F-QD-F system under consideration can be
written as 
\begin{eqnarray}
H &=&\sum_\sigma \varepsilon _dd_\sigma ^{+}d_\sigma +Ud_{\uparrow
}^{+}d_{\uparrow }d_{\downarrow }^{+}d_{\downarrow }+R(d_{\uparrow
}^{+}d_{\downarrow }+\text{H.c.})  \nonumber \\
&&+\sum_{k\alpha \in L,R\sigma }\epsilon _{k\alpha \sigma }a_{k\alpha \sigma
}^{+}a_{k\alpha \sigma }+\sum_{k\alpha \in L,R\sigma }\left[ V_{k\alpha
\sigma }a_{k\alpha \sigma }^{+}d_\alpha +\text{H.c.}\right]  \eqnum{1}
\end{eqnarray}
Here, the single particle energy $\varepsilon _d$ is double degenerate in
the spin index $\sigma $, and the interaction is included through the
Coulomb repulsion $U$. The first two terms in $H$ represent the correlated
level of the QD, the third term is used to describe potential spin-orbit
coupling which may cause the spin rotation of an electron while in the QD.
The spin-flip mechanisms that are relevant to the GaAs-based QD have been
studied recently\cite{Kha}. The fourth term describes the free magnetic
electrodes and the last term is the spin-dependent hybridization of the QD
to the magnetic electrodes. Model (1) has been employed to study tunnel
magnetoresistance in the sequential-tunneling regime\cite{Rud}.

Since the spin quantization axes in the electrodes are fixed by the internal
magnetization of the magnets, the electrons tunnel into a superposition of
spin-up and spin-down states. This coherence has to be taken into account
when calculating current\cite{Usaj}. Technically, we introduce a spin
rotation transformation $d_{\uparrow (\downarrow )}=(1/\sqrt{2})(c_{\uparrow
}\mp c_{\downarrow })$, in terms of which the dot Hamiltonian in Eq. (1) is
rewritten as $\sum_\sigma \varepsilon _{c\sigma }c_\sigma ^{+}c_\sigma
+Uc_{\uparrow }^{+}c_{\uparrow }c_{\downarrow }^{+}c_{\downarrow }$ with $%
\varepsilon _{c\sigma }=\varepsilon _d\pm R$ for up and down spins,
respectively. The current through the left electrode can be calculated from
the time evolution of the occupation number $N_L=\sum_{k\sigma }a_{kL\sigma
}^{+}a_{kL\sigma }$ for electrons in the left electrode using nonequilibrium
Green functions. The result is 
\begin{equation}
J=e\langle \stackrel{.}{N}_L\rangle =\frac{2e}{\hslash }\sum_k\int \frac{%
d\epsilon }{2\pi }\text{Tr}[{\bf V}_{kL}{\bf G}_{kL}^{<}(\epsilon )], 
\eqnum{2}
\end{equation}
where we have defined the tunneling amplitude matrix 
\begin{equation}
{\bf V}_{k\alpha }=\frac 1{\sqrt{2}}\left( 
\begin{array}{ll}
V_{k\alpha \uparrow } & V_{k\alpha \downarrow } \\ 
-V_{k\alpha \uparrow } & V_{k\alpha \downarrow }
\end{array}
\right) ,  \eqnum{3}
\end{equation}
$\alpha =L$, $R$, and lesser Green functions $[{\bf G}_{kL}^{<}(t)]_{\sigma
\sigma ^{\prime }}=i\langle a_{kL\sigma }^{+}(0)c_{\sigma ^{\prime
}}(t)\rangle $. Next, we use Dyson's equation to calculate the
non-equilibrium Green functions and express $J$ by the Green functions ${\bf %
G}_c$ of the dot as follows 
\begin{equation}
J=\frac e{\hslash }\int \frac{d\epsilon }{2\pi }\text{{\bf Tr}}\{{\bf \Gamma 
}^L(\epsilon )[i{\bf G}_c^{<}(\epsilon )+f_L(\epsilon ){\bf A(\epsilon )]}\},
\eqnum{4}
\end{equation}
where $[{\bf G}_c^{<}{\bf (}t{\bf )]}_{\sigma \sigma ^{\prime }}=i\langle
c_{\sigma ^{\prime }}^{+}(t)c_\sigma (0)\rangle $ is the matrix expression
for the lesser Green functions. ${\bf A(\epsilon )}=i[{\bf G}_c^R(\epsilon )-%
{\bf G}_c^A(\epsilon )]$ is the spectral function. $f_\alpha (\epsilon )$ is
the Fermi-distribution function in the $\alpha $ electrode, and 
\begin{equation}
{\bf \Gamma }^\alpha (\epsilon )=\frac 12\left( 
\begin{array}{ll}
\Gamma _{\uparrow }^\alpha (\epsilon )+\Gamma _{\downarrow }^\alpha
(\epsilon ) & \Gamma _{\downarrow }^\alpha (\epsilon )-\Gamma _{\uparrow
}^\alpha (\epsilon ) \\ 
\Gamma _{\downarrow }^\alpha (\epsilon )-\Gamma _{\uparrow }^\alpha
(\epsilon ) & \Gamma _{\uparrow }^\alpha (\epsilon )+\Gamma _{\downarrow
}^\alpha (\epsilon )
\end{array}
\right)  \eqnum{5}
\end{equation}
is the line-width matrix with $\Gamma _\sigma ^\alpha (\epsilon )=2\pi
\sum_{k\in \alpha }|V_{k\alpha \sigma }|^2\delta (\epsilon -\epsilon
_{k\alpha \sigma })$. The spin-dependence of $\Gamma _\sigma ^\alpha
(\epsilon )$ originates from the bulk magnetization of the electrodes.

In order to determine the retarded Green functions we choose the equation of
motion (EOM) method. Although this method does not provide quantitative
results in the Kondo regime, it gives the qualitative feature both in
equilibrium and nonequilibrium cases\cite{Mei1}. The method generates
higher-order Green functions, which have to be truncated to close the
equation. In the infinite-$U$ limit we obtain 
\begin{equation}
{\bf G}_c^R(\epsilon )=[\epsilon {\bf I}-\widehat{{\bf \varepsilon }}_c-{\bf %
\Sigma }_0^R(\epsilon )-{\bf \Sigma }_1^R(\epsilon )]^{-1}({\bf I}-{\bf n}_c)
\eqnum{6}
\end{equation}
where ${\bf I}$ is a $2\times 2$ unit matrix, $(\widehat{{\bf \varepsilon }}%
_c)_{\sigma \sigma ^{\prime }}=\delta _{\sigma \sigma ^{\prime }}\varepsilon
_{c\sigma }$, ${\bf \Sigma }_0^R=-i({\bf \Gamma }^L+{\bf \Gamma }^R)/2$ is
the self-energy matrix due to the tunneling coupling between the electrodes
and dot, ${\bf \Sigma }_1^R$ is due to the many-body correlation, with the
matrix elements $[{\bf \Sigma }_1^R(\epsilon )]_{\sigma \sigma
}=\sum_{k\alpha }[\frac{|V_{k\alpha \uparrow }|^2f(\epsilon _{k\alpha
\uparrow })}{\epsilon -\epsilon _{k\alpha \uparrow }-\varepsilon _{c\uparrow
}+\varepsilon _{c\downarrow }}+\frac{|V_{k\alpha \downarrow }|^2f(\epsilon
_{k\alpha \downarrow })}{\epsilon -\epsilon _{k\alpha \downarrow
}-\varepsilon _{c\uparrow }+\varepsilon _{c\downarrow }}]$, and $[{\bf %
\Sigma }_1^R(\epsilon )]_{\sigma \overline{\sigma }}=\sum_{k\alpha }[\frac{%
|V_{k\alpha \uparrow }|^2f(\epsilon _{k\alpha \uparrow })}{\epsilon
-\epsilon _{k\alpha \uparrow }}-\frac{|V_{k\alpha \downarrow }|^2f(\epsilon
_{k\alpha \downarrow })}{\epsilon -\epsilon _{k\alpha \downarrow }}]$, $(%
{\bf n}_c)_{\sigma \sigma ^{\prime }}=\delta _{\sigma \sigma ^{\prime
}}\langle c_{\overline{\sigma }}^{+}c_{\overline{\sigma }}\rangle $ must be
calculated self-consistently.

To solve the lesser Green functions ${\bf G}_c^{<}$ we generalize Ng's ansatz%
\cite{Ng2} to the present case. The interacting lesser and greater
self-energies are assumed to be of the form ${\bf \Sigma }^{<,>}={\bf \Sigma 
}_0^{<,>}{\bf B}$, where ${\bf B}$ is a matrix to be determined by the
condition ${\bf \Sigma }^{<}-{\bf \Sigma }^{>}={\bf \Sigma }^R-{\bf \Sigma }%
^A$. This ansatz is exact in the non-interacting limit ($U=0$) and
guarantees automatically the current conservation law. As a result one
obtains ${\bf \Sigma }^{<}={\bf \Sigma }_0^{<}[{\bf \Sigma }_0^R-{\bf \Sigma 
}_0^A]^{-1}[{\bf \Sigma }^R-{\bf \Sigma }^A]$. Using this ansatz, ${\bf G}%
_c^{<}$ can be obtained by Keldysh equation ${\bf G}_c^{<}={\bf G}_c^R$ $%
{\bf \Sigma }^{<}{\bf G}_c^A$. Substituting the expressions of the QD's
Green functions in Eq. (4), defining $\overline{{\bf \Sigma }}^{<}={\bf %
\Gamma }^R{\bf (\Gamma }^L{\bf +\Gamma }^R{\bf )}^{-1}({\bf \Sigma }^R-{\bf %
\Sigma }^A)$, and after a straightforward calculation, one obtains a compact
expression of the tunneling current 
\begin{equation}
J=\frac e{\hslash }\int \frac{d\epsilon }{2\pi }\text{Tr}[{\bf \Gamma }^L%
{\bf G}_c^R\overline{{\bf \Sigma }}^{<}{\bf G}_c^A][f_L(\epsilon
)-f_R(\epsilon )]  \eqnum{7}
\end{equation}
This expression generalizes the current formula in Ref. [16] to the
spin-dependent Anderson model with additional spin-flip relaxation and
allows one to describe the coherent spin transport through an interacting
quantum dot coupled to magnetic electrodes.

In the following calculations, for simplicity we neglect the energy
dependence in the tunneling matrix elements. The intrinsic linewidth of the
dot energies has a form $\Gamma _\sigma ^\alpha (\epsilon )=\Gamma _\sigma
^\alpha \theta (W-|\epsilon |)$ with the electrode band width $2W\gg \max
(k_BT$, $eV$, $\Gamma ^\alpha )$. We consider two magnetic configurations,
namely, parallel and antiparallel configurations. When the magnetic
electrodes are in parallel configuration, we assume that the spin-majority
electrons are up ($\sigma =\uparrow $) and the spin-minority electrons are
down ($\sigma =\downarrow $). We further assume that in the antiparallel
configuration the magnetization of the right electrode is reversed.
Therefore the spin dependence of the coherent transport can be conveniently
considered by introducing magnetic polarization factors $p_L$ and $p_R$ for
the left and right barriers, respectively. $\Gamma _{\uparrow (\downarrow
)}^L=\Gamma _0(1\pm p_L)$, $\Gamma _{\uparrow (\downarrow )}^R=\alpha \Gamma
_0(1\pm p_R)$ is for the parallel configuration, and $\Gamma _{\uparrow
(\downarrow )}^R=\alpha \Gamma _0(1\mp p_R)$ for the antiparallel
configuration. $\Gamma _0$ describes the coupling between the quantum dot
and the left electrode without internal magnetization and $\alpha $ denotes
tunnel asymmetry between the left and right barriers. In this work we assume
the symmetric barriers, i.e., $\alpha =1$, $p_L=p_R=p$.

The spin-resolved spectral densities are calculated via the relation $\rho
_{\uparrow (\downarrow )}(\epsilon )=-\frac 1\pi 
\mathop{\rm Im}%
\{({\bf G}_c^R)_{\uparrow \uparrow }+({\bf G}_c^R)_{\downarrow \downarrow
}\mp ({\bf G}_c^R)_{\uparrow \downarrow }\mp ({\bf G}_c^R)_{\downarrow
\uparrow }\}$. Figure 2 shows the behavior of the spectral densities in
parallel and antiparallel magnetic configurations, which will be used to
discuss the conductance results below. The Kondo resonance for each spin is
clearly manifested by a sharp peak at $\epsilon =0$ (the chemical potential
is set to be zero) in spectral densities for both magnetic configurations.
However, the peak shape is sensitive to the magnetic configurations of the
electrodes. As observed from Fig. 2(a), in parallel configuration the
excitation characteristics of the spectral densities are remarkably
different from the normal case in two prominent ways: (i) the Kondo
resonance for down spin is enhanced, while the up-spin resonance is
suppressed; (ii) interestingly, the broad single-particle resonances shift
away from the dot level $\varepsilon _d$, but in opposite directions for
different spins. The excitations for the down spin shifts towards higher
energy, while it shifts to lower energy for the up spin. This splitting is
due to the spin-dependence of the interacting self-energy matrix ${\bf %
\Sigma }_1^R(\epsilon )$, whose real part is different for the up and down
spins and sensitive to the values of the spin polarization factor $p$. From
the expression of ${\bf \Sigma }_1^R(\epsilon )$, the renormalized spin
levels $\widetilde{\varepsilon }_{d\sigma }$ are given by the
self-consistent equation 
\begin{eqnarray}
\widetilde{\varepsilon }_{d\sigma } &=&\varepsilon _d+\sum_{k\alpha }\frac{%
|V_{k\alpha \overline{\sigma }}|^2f(\epsilon _{k\alpha \overline{\sigma }})}{%
\widetilde{\varepsilon }_{d\sigma }-\epsilon _{k\alpha \overline{\sigma }}} 
\nonumber \\
&=&\varepsilon _d+\sum_\alpha \frac{\Gamma _{\overline{\sigma }}^\alpha }{%
2\pi }\{\ln \frac{2\pi k_BT}W+%
\mathop{\rm Re}%
\Psi [\frac 12-i\frac{\widetilde{\varepsilon }_d-\mu _\alpha }{2\pi k_BT}]\},
\eqnum{8}
\end{eqnarray}
where $\Psi $ is the digamma function. The result of Eq. (8) is shown in the
inset of Fig. 2 (a), where the dressed dot levels $\widetilde{\varepsilon }%
_{d\sigma }$ are plotted as a function of $p$. This spin splitting of the
dot levels in the interacting regime due to the magnetic properties of the
electrodes leads to essential changes in the transport properties (see
below). In addition, as $p$ increases, the spectral weight of $\rho
_{\downarrow }(\epsilon )$ goes up, while the spectral weight of $\rho
_{\uparrow }(\epsilon )$ is reduced. Thus a net magnetic moment is induced
by the magnetic coupling. In the case of antiparallel configuration, the
spectral densities of two spins are identical and the Kondo peak is not
influenced by the presence of magnetic polarization of the electrodes [see
Fig. 2(b)].

The level dressing and Kondo resonances in Fig. 2(a) are suppressed when
taking into account the spin-flip process. Remarkably, it shows in Fig. 3(a)
that a large spin-flip transition $R$ splits the original Kondo peak into 
{\it three} well-defined peaks. As seen from the expression of ${\bf \Sigma }%
_1^R(\epsilon )$, besides the peak at $\epsilon =0$, the additional two
peaks appear at $\epsilon =\pm R$, respectively. It is different from the
antiparallel case with symmetric barriers, in which, as in the normal case,
only the two Kondo peaks evolves from the presence of the spin-flip process
[see Fig. 3(b)].

The dramatic changes in the spectral densities when rotating magnetic
moments of the electrodes from parallel to antiparallel configuration
suggests substantial different transport properties for these two
configurations. Figure 4 shows the linear response conductance $G$ as a
function of $\varepsilon _d$, which can be tuned via the external gate
voltage, for different temperatures. In the antiparallel configuration, as
observed from Fig. 4(a), the temperature dependence of the conductance is
similar to the normal case although with a lower amplitude: the Kondo
resonance broadens the conductance peak and saturates the peak amplitude at
low temperatures. In addition, the peak shape remains nearly symmetric over
a broad range of temperatures. In the parallel configuration, however, the
conductance peak becomes asymmetric with decreasing temperature as shown in
Fig. 4(b). This asymmetry is even more pronounced in the spin-resolved
conductance $G_\sigma $ [see the inset of Fig. 4(b)]. The peak splitting in $%
G_{\uparrow }$ and $G_{\downarrow }$ is in the same manner as shown in Fig.
2 due to the magnetic dressing of spin levels. Thus their superposition
results in a double-peak structure at low temperatures and for large $p$ as
shown in Fig. 4(b) (doted line). The main peak with larger amplitude is
dominated by the up-spin resonance, while the other peak with lower
amplitude nearly comes from the down-spin resonance. This novel spin
filtering effect is fully caused by the interplay between the dot
correlation and the magnetic coupling. When spin-flip transition is
included, as shown in Fig. 4(c), the splitting of the conductance is
suppressed.

To describe the dramatic change of $G$ when the magnetic moments in the
electrodes are rotated from antiparallel to parallel configuration, we
define the linear magnetoresistance as $MR=(G_p-G_{ap})/G_{ap}$, where $G_p$
and $G_{ap}$ are the linear conductance in the parallel and the antiparallel
configurations, respectively. Figure 5 shows the magnetoresistance as a
function of $\varepsilon _d$ for different temperatures. One can see that
when the dot level is far from resonance, the magnetoresistances approach to
the same value for different temperatures, coinciding with the
non-interacting case. At low temperatures, however, dramatic changes occur
in the resonant tunneling regime. The magnetoresistance develops into a dip
with a negative value. This means that the linear magnetoresistance
significantly decreases and even changes its sign at low gate voltages. In
the empty orbital regime where the dot level is higher than the chemical
potential of the electrodes, the magnetoresistance may be enhanced to a
value as large as 160\%.

In summary, using Anderson model, it is shown that the magnetic moment
arrangement in the electrodes plays an essential role in spin-dependent
transport of a F-QD-F system in the interacting regime. For parallel
magnetic configuration, the Kondo resonance and QD energy levels can be
controlled by the magnetic polarization in the electrodes. Consequently, the
linear conductance appears as a spin-resolved double-peak structure and a
net magnetic moment emerges in the QD. The spin-flip process in the QD
results in a splitting of the Kondo peak into three peaks. We expect these
results are useful in exploiting the role of electronic correlation in
spintronics.

This work is supported by CNSF under Grant No. 69625608 and by US-DOE.

{\it Note added}.---After the work was completed, we noticed that Kondo
problem in TMR has also been considered by Sergueev {\it et al}.\cite{Guo}.
However, they did not discuss spin splitting of the conductance and
spin-flip effects, which compose essential points in this paper.

{\Large Figure captions}

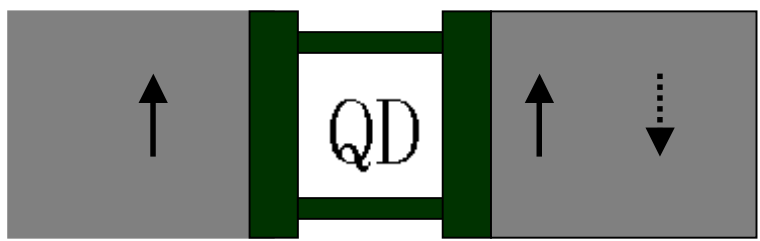
Fig. 1. Schematic plot of the F-QD-F system considered in this work.

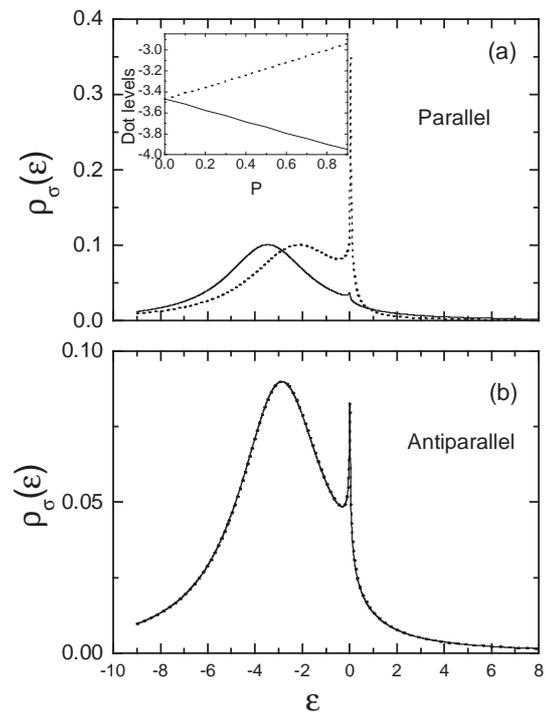
Fig. 2. $\rho _{\uparrow }(\epsilon )$ (solid line) and $\rho _{\downarrow
}(\epsilon )$ (dotted line) as a function of $\epsilon $ in the (a)
parallel, and (b) antiparallel magnetic configurations for $p=0.5$, $%
k_BT=0.02\Gamma _0,$ $\varepsilon _d=-4\Gamma _0$, and $R=0$. Inset in (a)
indicates spin splitting of the dot levels as a function of $p$.

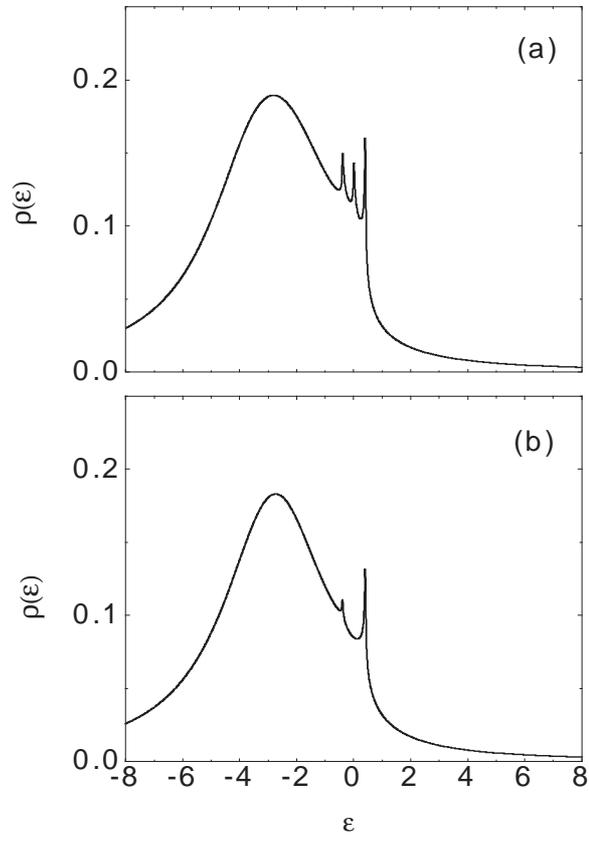
Fig. 3. $\rho (\epsilon )$ as a function of $\epsilon $ in the (a) parallel,
and (b) antiparallel magnetic configurations with $R=0.2\Gamma _0$. Other
parameters are the same as in Fig. 2.

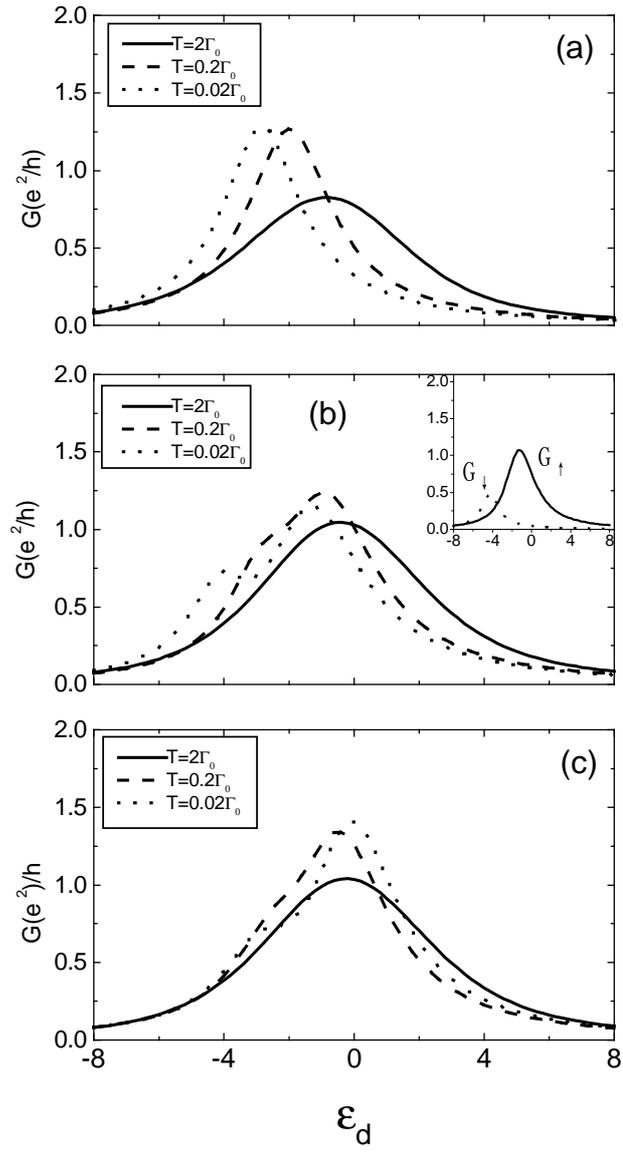
Fig. 4. Linear conductance $G$ as a function of $\varepsilon _d$ in the (a)
antiparallel, and (b) parallel magnetic configurations in the absence of
spin-flip process with $p=0.5$. Inset indicates spin-resolved conductance $%
G_{\uparrow }$ and $G_{\downarrow }$. The effect of of intradot
spin-flipping on the parallel conductance is shown in (c).

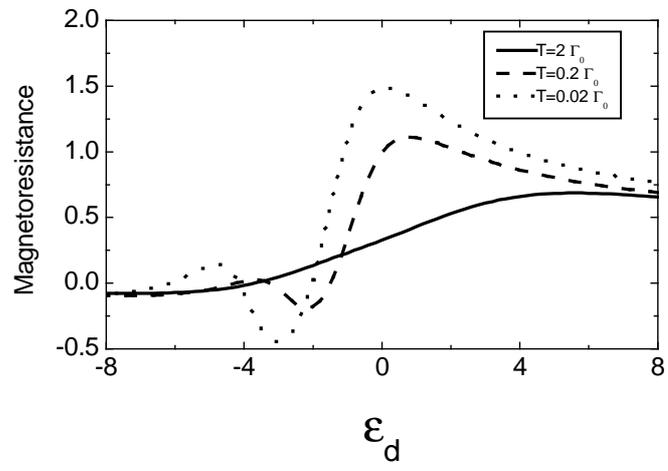
Fig. 5. Linear magnetoresistance as a function of $\varepsilon _d$ for
several values of temperature. Other parameters are the same as in Fig. 4.

\end{document}